\documentstyle[hyperref,11pt]{article}
\hypersetup{colorlinks,pagebackref,bookmarksopen,breaklinks,pdftitle={Third-order autoparallel spin dynamics},pdfdisplaydoctitle=true}
\makeatletter
\def\ps@firstpage{\ps@plain
  \def\@oddfoot{\normalfont\scriptsize \hfil\thepage\hfil}
  \def\@oddhead{\article@logo\hss}}
\def\article@logo{%
  \vbox to\headheight{%
    \@parboxrestore \@logofont
    \noindent
\href{http://www.univie.ac.at/EMIS/proceedings/7ICDGA/V/}{Differential Geometry and Its Applicatons},
    \newline Proc. Conf., Opava (Czech Republic), August 27--31, 2001.
    \newline Silesian University, Opava, 2001, 447--459.
    \par\vss
  }}
\def\@logofont{\fontsize{8}{9.6\p@}\selectfont}
\makeatother
\font\frak=eufm10  scaled \magstephalf 
\font\ssb=cmssbx10 scaled \magstephalf 
\font\bbold=msbm10    scaled \magstephalf 

\def\bcdot{\mbox{\boldmath$\cdot$}}
\def\bmat#1{\mbox{\boldmath$#1$}}

\newbox\Elocity\setbox\Elocity=\hbox{\ssb v}
\newbox\PrimE\setbox\PrimE=\hbox{{\ssb v}$^{\prime}$}
\newbox\piN\setbox\piN=\hbox{\ssb s}
\catcode`V=13\def V{\copy\Elocity}\catcode86=11
\catcode`W=13\def W{\copy\PrimE}\catcode87=11
\catcode`S=13\def S{\copy\piN}\catcode83=11
\def\b{\begingroup\catcode`S=13 \catcode`V=13 \catcode`W=13}
\def\e{\endgroup}

\newcommand{\SS}{\scriptscriptstyle}
\newtheorem{prop}{Proposition}
\newtheorem{rem}{Remark}
\def\q{{\bmat{^2}}}
\def\ldot{{\;.\;}}
\def\bld{\mbox{\boldmath$\;.\;$}}
\def\RN#1{\uppercase\expandafter{\romannumeral #1}}
\def\bb#1{\mbox{\bbold #1}}
\def\R{\mbox{\bbold R}}
\def\C{C^r(p,M)}
\def\T{T^r_pM}
\def\TT{T^{2r}_pM}
\def\J{J^r(\R^p,M)}
\def\JJ{J^{2r}(\R^p,M)}
\def\JR{J^r(\R^p,\R^q)}
\def\JRf{J^r(\R^p\times\R^q)}
\def\Jf{J^r(f,id)}
\def\JJf{J^{2r}(f,id)}
\def\j{j^r}
\def\jj{j^{2r}}
\def\js{j^r\sigma}
\def\p{p_2(r)}
\def\pp{p_2(2r)}
\def\d{\bmat\delta}
\def\dt{\bmat\delta^T}
\def\L{{\cal L}}
\def\l{{\cal L}_{\SS0}}
\def\M{{\rm M}}
\def\N{{\rm N}}
\def\Y{{Y^r}}
\def\V{T(Y^r/\R^p)}
\def\Vd{T^*(Y^r/\R^p)}
\def\O#1#2{\Omega^{#1,#2}_r}
\def\OO#1#2{\Omega^{#1,#2}_{2r}}
\def\fp{\mbox{\frak p}}
\renewcommand{\thefootnote}{ }
\title{Autoparallel variational
description of the free relativistic top third order dynamics}
\author{Roman Matsyuk\\Institute for Applied Problems in Mechanics and
Mathematics\\ 15~Dudayev~Str., 290005 L\kern-1pt'viv, Ukraine
\\romko.b.m@gmail.com, matsyuk@lms.lviv.ua
\\\url{http://www.iapmm.lviv.ua/12/eng/files/st_files/matsyuk.htm}}
\begin{document}
\maketitle
\thispagestyle{firstpage}
\renewcommand{\thefootnote}{}
 \footnotetext{MS
2000 classification: 83C10, 70H40, 70H50, 70G45, 58E30}
\footnotetext{The paper is in final form and  will not be
published elsewhere} \footnotetext{Research supported by Grants
MSM:J10/98:192400002 of the Ministry of Education, Youth and
Sports, and GACR 201/00/0724 of the Grant Agency of the Czech
Republic.}
\renewcommand{\thefootnote}{\arabic{footnote}}\setcounter{footnote}{0}
\begin{abstract}
A second order variational description of the
autoparallel curves of some differential-geometric connection for the third
order Mathisson's ``new mechanics'' of a relativistic free spinning particle
is suggested starting from general requirements of invariance and
``variationality''.
\end{abstract}
\section{Introduction}
In 1937 Mathisson in the article named ``Neue Mechanik
materieller Sy\-s\-te\-me''\cite{Math} introduced a third order differential
equation to describe the motion of quasi-classical relativistic particle with
inner angular momentum given by a skewsymmetric tensor $S^{\alpha\beta}$:
\begin{equation}
\label{Math}
m_{0}\frac{Du^{\alpha}}{d\tau}=S^{\alpha\beta}\frac{D^2u_{\beta}}{d\tau^2}
-{1\over2}R^{\alpha}_{\beta\gamma\delta}u^{\beta}S^{\gamma\delta},
\end{equation}
where the velocity four-vector $u^{\alpha}, \alpha\in\overline{(0,3)}$ is
subject to the usual constraint $u^{\alpha}u_{\alpha}=1$. Equation
(\ref{Math}) in fact was considered by Mathisson under the assumption of
later well known ``Pirani auxiliary condition''
\begin{equation}
\label{Pir1}
u_{\beta}S^{\alpha\beta}=0,
\end{equation}
which fixes one possible way of choosing the point of reference
within the tube of world lines followed by different points of an
extended object with dipole angular momentum $S^{\alpha\beta}$.
This way or that, one may pose the following question: what
geometry is best suited for the description of physical particles
with complicated internal structure? In presence of the
gravitational field such geometry of course will incorporate the
curvature tensor, but the other question arises then to invent a
local model for such a future geometry. And with this approach in
mind, we start with the pseudo-Euclidean space, endowed not only
with the usual structure of geodesic straight lines, but also with
some other structure, the autoparallel curves of which would
satisfy also the unparametrized version of Mathisson's equation
(\ref{Math}) with zero curvature tensor
$R^{\alpha}_{\beta\gamma\delta}$.

The constraint (\ref{Pir1}) suggests the idea to introduce the spin
four-vector
$$s_{\delta}=\frac{1}{2\|\bmat
u\|}\varepsilon_{\alpha\beta\gamma\delta}u^\alpha
S^{\beta\gamma},$$
and it was proved in \cite{Thesis} and
published in \cite{DAN} that in terms of this spin vector the
Mathisson equation (\ref{Math}) is equivalent to the following one
(we put $R^{\alpha}_{\beta\gamma\delta}=0$):
\begin{equation}
\label{DAN}
\varepsilon_{\alpha\beta\gamma\delta}\ddot u^\beta u^\gamma s^\delta
-3\frac{\dot u_\beta u^\beta}{\|\bmat u\|^2}\,
    \varepsilon_{\alpha\beta\gamma\delta}\dot u^\beta u^\gamma s^\delta
-m_0\left(\|\bmat u\|^2 \dot u_\alpha - \dot u_\beta u^\beta u_\alpha \right)
=0,
\end{equation}
subject to the constraint
\begin{equation}
\label{Pir2}
s_\alpha u^\alpha = 0.
\end{equation}

And we recall that spin four-vector $\bmat s$ is a constant vector along the
word line of the particle as long as no gravitational field is considered.

The equation (\ref{DAN}) does not change under arbitrary reparametrizations of
the world line (i.e. under arbitrary local transformations of the independent
variable $\tau$, the parameter, and because of that it is often said that the
equation is presented in homogeneous form, or that it is
parameter-independent.

Now, we set the following twofold task: 1)~invent a variational
description for equation~(\ref{DAN}); 2)~try to add some parametrization
to equation~(\ref{DAN}) in such a way, that the
(parametrized) autoparallel curves of the corresponding second order connection
would also satisfy (\ref{DAN}) everywhere on the constraint
submanifold (\ref{Pir2}).

\section{Variationality}
As far as we are interested in
the parameter-homogeneous form of a variational third-order
equation that should be equivalent to the equation (\ref{DAN}), and
also as far as we intend to impose pseudo-Euclidean symmetry, it
is convenient to work in the variables $u^0=1$, $v^i=u^i$,
$i\in\overline{(1,3)}$, $x^0=t$, that is to pass to the manifold
of $r$-th order contact elements in the manifold $M=\{t,x^i\}$. We
note that pseudo-Euclidean transformations permute the variables
$t$ and $x^i$.

Let, in general, $\T$ denote the bundle of Ehresmann $p$-velocities
of order $r$ to manifold $M$ and let $\C$ denote the manifold of
$r$-th order contact elements of $p$-dimensional submanifolds in
$p+q$ dimensional manifold $M$. The group $Gl^r(p)$ of invertible
jets from $\R^p$  to $\R^p$ which both start and terminate at
$0\in\R^p$, acts on the right upon the manifold $\T$ by jet
composition rule. This action, as we shall see, is in charge of
parameter (independent variable) transformations of the velocities
from $\T$ and hence governs the transformations of a variational
equation in parametric form.  The generators are the generalized
Liouville fields
\begin{equation}
\label{Zeta}
\bmat\zeta^{\M}_{n}=\sum^{r-|\M|}_{|\N|=0}{|\M|+|\N|\choose
|\M|}u^\alpha_{\N+1_n}\frac{\partial}{\partial u^\alpha_{\N+\M}}, \quad
1\leq|\M|\leq r,
\end{equation}
where, as common, multi-indexes $\M=(\mu_1,\ldots,\mu_p)$ and
$\N=(\nu_1,\ldots,\nu_p)$ both belong to $\bb N^p$ with the length  defined
by $|\N|=\nu_1+\ldots+\nu_p$, and the multi-index $1_n$ corresponds to partial
differentiation along the direction of the $n$-th independent variable
$\tau^n$, $n\in\overline{(1,p)}$. In future we shall abuse the notation
$u^\alpha_0$ in place of $x^\alpha$, $x^\alpha=t^\alpha$ if $\alpha\leq p$.
The zero section of $\T$ is well defined and we have the quotient projection with respect to the above mentioned action
\begin{equation}
\label{pcont} \wp:\T\backslash\{0\}\to\C.
\end{equation}

On the manifold $M$ we shall define a variational problem,
invariant under the action of pseudo-Euclidean group on $M$. A
Lagrangian will mean a semi-basic with respect to $M$ local
$p$-form defined on $\C$, and two such forms will be recognized as
equivalent if in common domain their difference belongs to the
ideal, generated by contact forms. As our considerations on $\C$
are local and infinitesimal, we shall profit from the local
isomorphism
$$\C\approx\JR.$$
And further on, let us recall the
isomorphism $\JR\approx\JRf$, where the right hand side means the
bundle of jets of cross sections of the fibration
$\R^{p+q}\to\R^p$. From among the equivalent Lagrangians on $\C$
it is always possible to fix the unique representative, semi-basic
with respect to $\R^p$ in this local representation.

Let us introduce the notation $v^i_\Omega$, with
$\Omega=(\omega_1,\ldots,\omega_p)$, for the canonical coordinates
in $\JR$ and let $(t^w,x^i)$, $w\in\overline{(1,p)}$,
$i\in\overline{(1,q)}$, be the corresponding local coordinates in
$M$. One would like to pull the variational problem posed on $\C$,
back to the manifold $\T$ in the temptation to obtain some
variational equation in the parameter-homogeneous form on $M$, and
in case $p=1$ to construct then a kind of higher-order connection
on some $T^kM$, $k<2r$, in such a way, that the autoparallel
curves of this connection would prescribe some parametrization to
the unparametrized integral submanifolds of the initial
parameter-independent variational problem. But the pull-back of a
one-form is again one-form, and what we need is a local Lagrange
{\it function} on $\T$, not a form. The way out is to consider the
manifold $\T$ as a rudiment of the parameter-extended space $\J$
in the following way.

First, recall the isomorphism $\J\approx\R^p\times\J(0)$, given by the
correspondence $\js(\tau)\to(\tau,\j(\sigma\circ\delta_\tau)(0))$, where
$\delta_\tau$ is the translation by $\tau$ in $\R$. Then notice that $\J(0)$
is exactly the definition of $\T$ and apply the projection onto the second
factor,
\begin{equation}
\label{pr2}
\J\approx\R^p\times\T\stackrel{\textstyle\p}{\longrightarrow}\T.
\end{equation}

Now, the idea is to pull a variational problem from the manifold $\C$ back to
the manifold $\J$ and then to find on $\J$ an equivalent Lagrangian of the
form $\l d\tau^1\wedge\cdots\wedge d\tau^p$. The function $\l$ will then in
fact be defined on the space $\T$. To make our consideration precise, let us
recall some calculus on $\J$.

\subsection{Lagrange differential}
Let us introduce an abridged notation $\Y=\J$ and, of course, $Y$ will stand in
place of $\R^p\times M$. Also let $\O\bcdot\bcdot =\sum\O h v$
denote the module of semi-basic with respect to $\R^p$ differential forms on
$\Y$ with values in the dual $\Vd$ to the bundle $\V$ of $\R^p$-vertical
tangent vectors to $\Y$; $h$ and $v$ mean the corresponding degrees in
the bigraded module
$$ \O h v \approx
Sec\left(\wedge^v\Vd\otimes_\Y\wedge^hT^*\R^p\right). $$
It is not
our goal here to present any definition of the Euler-Lagrange
differential $\d$ (see \cite{Kolar} or \cite{Studii}). We merely
recall that it is possible to interpret the operator $\d$ as one
acting from $\O h v$ to $\OO h {v+1}$ so that for any $\lambda\in
\O p 0$ the result of applying $\d$ belongs to $\OO p 1$, and in
fact $\d\lambda$ is a semi-basic $p$-form taking values in
$T^*(Y/\R^p)$ alone. Its components in $T^*(Y/\R^p)$ along some
local coordinates $\{x^\alpha\}$ in $M$ are the classical
Euler-Lagrange expressions. Let us identify the fiber bundle
$\wedge\Vd$ with the reciprocal image of $\wedge T^*(\T)$ along
the projection (\ref{pr2}). We think of the algebra $\Omega(\T)$
of differential forms on $\T$ as of $\Omega^0(\T)$-subalgebra of
$\O 0\bcdot$, the inclusion being defined by the reciprocal image
construction along $\p$. The operator $\d$ takes $\Omega(\T)$ into
the $\Omega^0(\TT)$-subalgebra $\Omega(\TT)$ of $\OO 0 \bcdot$. We
denote the restriction of the operator $\d$ to the algebra
$\Omega(\T)$ by $\dt$.

Now, consider some Lagrangian
\begin{equation}
\label{lambda}
\lambda=\l d^p\tau\in\O p0,
\end{equation}
where $d^p\tau$ stands for the $p$-fold exterior product
$d\tau^1\wedge\cdots\wedge d\tau^p$, and, in general, function
$\l$ may depend on $\tau\in\R^p$. We say that such a Lagrangian
defines a variational problem in extended parametric form. In this
case,
\begin{equation}
\label{epszero}
\d\lambda=\bmat\varepsilon_{\SS0}\otimes d^p\tau\in \OO p1, \qquad
\mbox{where}\qquad\bmat\varepsilon_{\SS0}=\d\l.
\end{equation}

Let
\begin{equation}
\label{tpr}
p^r:\T\to M
\end{equation}
denote the standard projection. We observe that essentially
$\bmat\varepsilon_{\SS0}$ is a cross-section of the induced bundle
$\pp^*{p^{2r}}^*T^*M$.

Let $\upsilon$ be the graph of a local immersion $\sigma:\R^p\to M$. Recall
that by the definition of the action of pull-backs  on vector bundle
valued differential forms,
$$
\jj\upsilon^*\d\lambda\in Sec\left(\sigma^*T^*M\otimes\wedge^pT^*\R^p\right)
\qquad\mbox{and}\qquad
\jj\upsilon^*\d\lambda=\left(\d\l\circ\jj\sigma\right)\otimes d^p\tau,
$$
where $\jj\upsilon$ denotes the prolongation of the cross-section
$\upsilon$ and $\jj\sigma$ is the essential component of the cross-section
$\jj\upsilon$. Thus the Euler-Lagrange equations appear to have two
equivalent guises:
\begin{equation}
\label{guises}
\jj\upsilon^*\d\lambda=0 \qquad\mbox{or}\qquad \d\l\circ\jj(\sigma)=0.
\end{equation}

Let us assume that the Lagrange function $\l$ does not depend on
parameter $\tau\in \R^p$: $\l=\p^*\L$. Concider the second component
$\partial_{2r}\sigma$ of the jet $\jj\sigma$ under the projection
$\pp:\JJ\to \TT$. The Euler-Lagrange equations take the shape
\begin{equation}
\label{Tequ}
\left(\dt\L\right)\circ\partial_{2r}\sigma=0.
\end{equation}

\subsection{Parametric invariance}
Do introduce an arbitrary local change of parameter $\R^p\to\R^p$ and let us see
how it effects a variational problem in extended parametric form on the
fiber manifold $\pi:\J\to\R^p$, given by (\ref{lambda}). The standard
prolongation of the pair of morphisms $(f,id):\R^p\times M\to\R^p\times M$ is
denoted by $\Jf:f^*\J\to\J$ and is defined by the property
\begin{equation}
\label{prol}
\Jf\circ (\pi^*f)^{-1}\circ\js\circ f = \j(\sigma\circ f)
\end{equation}
for arbitrary jet $\js\in\J$, and we mention that in standard
functorial notations morphism $\pi^*f:f^*\J\to\J$ is a bijection
as long as the mapping $f$ is a diffeomorphism. Let $(W,\sigma)$
be a pair consisting of a compact set $W$ in $\R^p$ and of a
mapping $\sigma$ from $W$ into $M$. Diffeomorphism $f$ acts upon
such pairs by means of the rule
$f:(W,\sigma)\mapsto(f^{-1}W,\sigma\circ f)$. Let $S$ be a
function, defined for each pair $(W,\sigma)$ by means of
$S:(W,\sigma)\mapsto \int_{W}\js^*\lambda$. We demand that the
function $S$ be equivariant with respect to the action of $f$,
that is,
\begin{equation}
\label{equiv}
S\circ f=S,
\end{equation}
and in this case the variational problem is called a parameter-invariant
one.  By (\ref{prol}) and by the well known change of variables formula,
$$
S(W,\sigma)=\int_{f^{-1}W}f^*\js^*\lambda,
$$
we obtain:
\begin{eqnarray*}
\lefteqn{(S\circ f)(W,\sigma)=S(f^{-1}W,\sigma\circ
f)=\int_{f^{-1}W}(\j(\sigma\circ f))^*\lambda=}\\ &
&\int_{f^{-1}W}f^*\js^*((\pi^*f)^{-1})^*(\Jf)^*\lambda=S\left[W,((\pi^*f)^{-1})^*(\Jf)^*\lambda\right].
\end{eqnarray*}
Now the parametric invariance (\ref{equiv}) means that
\begin{equation}
\label{par}
\Jf^*\lambda=(\pi^*f)^*\lambda.
\end{equation}
The identification (\ref{pr2}) implies that
$f^*\J\approx\R^p\times\T$ and $\pi^*f=(f\times id)$, thus
(\ref{par}) takes the form
\begin{equation}
\label{par1}
\l\circ\Jf=\l\circ(f\times id).\det\frac{\partial f}{\partial\tau}.
\end{equation}
The infinitesimal analogue of (\ref{par1}) reads
$$
\left\langle\bmat\zeta^r,{\bf d}_{\pi}\l\right\rangle=\bmat\zeta(\l)+\l{\rm
tr}\frac{\partial\bmat\zeta}{\partial\tau},
$$
where $\bmat\zeta$ generates some local flow on $\R$,
$\bmat\zeta^r$ denotes the standard prolongation of $\bmat\zeta$ to the space $\J$, and
${\bf d}_\pi$ is the fiber differential along fibers of $\pi$. As far as
$\frac{\partial\bmat\zeta}{\partial\tau}$ is an arbitrary matrix, we conclude
that $\l$ must not depend on $\tau$ (put
$\frac{\partial\bmat\zeta}{\partial\tau}=0$) and thus essentially is defined,
and may be thought of, as some function on $\T$ alone: $\l=\p^*\L$.
The calculation of $\bmat\zeta^r$ according to the standard procedure
\cite{Ovsian} ultimates in Zermelo-G\'eh\'eniau's conditions
\begin{equation}
\label{Zermelo}
\bmat\zeta^{\M}_n(\L)=\delta^{\M}_{1_n}\L,
\end{equation}
where fields $\bmat\zeta^\M_n$ are given by (\ref{Zeta}).

It is well known, that an invariance of Lagrangian $\lambda$ implies the
invariance of the corresponding differential form $\d\lambda$ (see
\cite{KolarLie} for technical details). In our notations,
$$ \JJf^*\d\lambda=(f\times id)^*\d\lambda, $$
and in terms of
projection (\ref{pr2}) it gives for
$\bmat\varepsilon_{\SS0}=\pp^*\dt\L$ as defined in
(\ref{epszero}),
$$
\JJf^*\pp^*\dt\L=(\pp^*\dt\L).\det\frac{\partial
f}{\partial\tau}.  $$
The infinitesimal analogue in terms of fiber derivative
${\bf d}_{p^{2r}}$ with respect to the fibration $p^{2r}:\TT\to M$ reads
$$
\left\langle\bmat\zeta^{2r},\pp^*{\bf d}_{p^{2r}}\dt\L \right\rangle
=(\pp^*\dt\L).{\rm tr}\frac{\partial\bmat\zeta}{\partial\tau},
$$
and again in course of the arbitrariness of $\bmat\zeta$ we come up to the
following formulation of parametric invariance of the Euler-Lagrange form
$\bmat\varepsilon_{\SS0}=\pp^*\dt\L$:
\begin{equation}
\label{epsZermelo}
\left\langle\bmat\zeta^{\M}_n,{\bf d}_{p^{2r}}\dt\L\right\rangle =\delta^{\M}_{1_n}\dt\L.
\end{equation}

\subsection{Transition from $\C$ to parameter-homogeneous form}
Projection (\ref{pcont}) in local coordinates is given by the following
formula, which may be deduced from general reflections on the
subject of transformation rules for derivatives \cite{Kaw}
$$
u^i_{n_1\ldots n_r}
=\sum^r_{k=1}{\bf P}^{w_1\ldots w_k}_{n_1\ldots n_r}\wp^i_{w_1\ldots w_k},
$$
where we put $\wp^i_{w_1\ldots w_k}=v^i_\Omega\circ\wp$, $v^i_\Omega$ being
the coordinates in $\C$ which coincide with $u^i_\N$ for $\N=\Omega$ when
$u^w_\N=\delta^{1_w}_\N$, and in the multi-index
$\Omega=(\omega_1\ldots\omega_p)$ of length $k$ each $\omega_w$ denotes
the number of repetitions of $w$ in the sequence $(w_1\ldots w_k)$.
The matrix $\bf P$ is calculated according to the formula:
\begin{eqnarray*}
\lefteqn{{\bf P}^{w_1\ldots w_k}_{n_1\ldots n_r}=
\sum_{1\leq r_1\leq\cdots\leq r_k\leq r\atop r_1+\cdots+r_k=r}
\frac{r!}{r_1!\cdots r_k!\rho_1!\cdots\rho_{r-k+1}!}}\\
&&\cdot\; u^{(w_1}_{(n_1\ldots n_{r_1}}u^{\mathstrut w_2}_{\mathstrut n_{r_1+1}\ldots
n_{r_1+r_2}}\cdots u^{w_k)}_{n_{r_1+r_2+\cdots+r_{k-1}+1}\ldots
n_{r_1+r_2+\cdots+r_k)}},
\end{eqnarray*}
where each $\rho_k$ means the
number of repetitions of $k$ in the sequence $(r_1\ldots r_k)$ and
parentheses denote the symmetrization procedure.

We now proceed further in the realization of our main goal: to
represent a variational problem, initially posed on the contact
manifold $\C$, by means of some pa\-ra\-me\-ter-ho\-mo\-ge\-neous
form of an equivalent variational problem, this time on the
manifold $\T$. Let be given in some local chart of $\C$ an
$\R^p$-semi-basic representative
\begin{equation}
\label{caplambda}
\Lambda=Ld^pt, \quad d^pt=dt^1\wedge\cdots\wedge dt^p
\end{equation}
of a class of equivalent Lagrangians (see \cite{Sym}). The
pull-back of $\Lambda$ along the total projection $\fp=\wp\circ\p$
equals $L\circ\fp
\ldot d^pt$. Let us decompose the $p$-form $d^pt$ with respect to
the basis, constituted by the $p$-form $d^p\tau$ and by the forms
$$
(d^p\tau)^{\alpha_1\ldots\alpha_l}_{n_1\ldots n_l}
=\vartheta^{\alpha_1}\wedge\cdots\vartheta^{\alpha_l}
\wedge\frac{\partial}{\partial
\tau^{n_1}}\rfloor\cdots\rfloor\frac{\partial}{\partial \tau^{n_l}}\rfloor
d^p\tau,
$$
$1\leq l\leq p$, $1\leq n_1<\cdots<n_l\leq p$, $1\leq
\alpha_1<\cdots<\alpha_l\leq p+q$, where
$\vartheta^\alpha=dx^\alpha-u^\alpha_n d\tau^n$ are the first order contact
forms on the manifold $\J$ and $\tau\in\R^p$. In fact, only terms with
$d^p\tau$ and $(d^p\tau)^{w_1\ldots w_l}_{n_1\ldots n_l}$, $1\leq
w_1<\ldots<w_l\leq p$ survive in this decomposition, and we obtain
$$
d^pt=\det{\bf U}.d^p\tau+\sum^p_{l=1}\sum_{1\leq n_1<\cdots<n_l\leq
p\atop 1\leq w_1<\cdots<w_l\leq p} \overline{{\bf U}}^{n_1\ldots n_l}_{w_1\ldots
w_l}(d^p\tau)^{w_1\ldots w_l}_{n_1\ldots n_l},
$$
where  $\overline{{\bf
U}}^{n_1\ldots n_l}_{w_1\ldots w_l}$ denotes the algebraic adjunct of the
minor ${\bf U}_{n_1\ldots n_l}^{w_1\ldots w_l}$ in the matrix ${\bf
U}=(u^w_n)$.

Consider for a moment another local chart $(u'^i,x'^i,t'^i)$ of
the manifold $\C$, denote by $\phi_C$ the corresponding transition
function and let $\Lambda'=L'd^pt'$ be such a representative, that
$\phi_C^*\Lambda'-\Lambda$ belongs to the ideal, generated by
differential forms
$$
(d^pt)^{i_1\ldots i_l}_{w_1\ldots w_l}=\\
\theta^{i_1}\wedge\cdots\theta^{i_l}
\wedge\frac{\partial}{\partial
t^{w_1}}\rfloor\cdots\rfloor\frac{\partial}{\partial t^{w_l}}\rfloor d^pt,
$$
$1\leq l\leq p$, $1\leq w_1<\cdots<w_l\leq p$, $1\leq
i_1<\cdots<i_l\leq q$, where $\theta^i=dx^i-v^i_w dt^w$ are the
first order contact forms on the manifold $\JR$ and $t\in\R^p$.
The pull back operation preserves the corresponding contact ideal
\cite{Sym}: $\fp^*\theta^i=\vartheta^i-\wp^i_w\vartheta^w$ as
well, as the coherent transition function $\varphi_J$ in the
manifold $\J$ does, and it may be proved that the difference
$\varphi_J^*(L'\circ\fp\ldot d^pt')-L\circ\fp\ldot d^pt$ belongs
to the contact ideal on $\J$. Hence our considerations are
intrinsic.

Let us recall the notations (\ref{lambda}, \ref{epszero},
\ref{guises}, \ref{Tequ}), and introduce the shortcut notation
$
\wp_\phi=\phi_C\circ\wp:\T\to\JR,
$
same for each $r$.

\begin{prop}
\label{P1}
Let $\L=(L\circ\wp_\phi).\det{\bf U}$. The equations $(\d
L)\circ\wp_\phi\circ\partial_{2r}\sigma=0$ and
$\dt(\L)\circ\partial_{2r}\sigma=0$ are equivalent.
\end{prop}
The Lagrange function $\L$ and the corresponding differential form $\dt\L$
obviously satisfy Zermelo-G\'eh\'eniau's  conditions (\ref{Zermelo}) and
(\ref{epsZermelo}).

\begin{rem}{\rm We strive to give an (in fact trivial) algorithm for
building up a Lagrange function and the corresponding
Euler-Lagrange equations in parameter-homogeneous form directly
from solutions of an inverse variational problem on contact
manifold $\C$. But treating this latter problem, especially in the
aspects of equivalence and symmetry of differential equations,
appears to be more convenient in terms of the {\it Lepagean
equivalents} \cite{Krup} (see \cite{Sym}, \cite{DGA95},
\cite{condenced}). So it would be of interest to translate the
reparametrization technique, presented in this paragraph, directly into the
language of Lepagean differential forms theory.}
\end{rem}

\subsection{Third order equations with pseudo-Euclidean symmetry}
In case of the system of ordinary differential Euler-Lagrange
equations (we follow the tradition of calling them {\it
Euler-Poisson equations}) the vector-valued differential form
$\d\Lambda$ of $\Lambda$ as in (\ref{caplambda}), takes the shape
\begin{equation}
\label{E} \d\Lambda=E_idx^i\otimes dt,
\end{equation}
where $E_{i}$ are the Euler-Poisson expressions. We call the
problem of finding Euler-Lagrange equations with prescribed
symmetry and of prescribed order, the invariant inverse problem of
that order in the calculus of variations. In case of third order
Euler-Poisson equations with pseudo-Euclidean symmetry in
four-dimensional space, one solution was found in \cite{Thesis}
and announced in \cite{DAN}. It is essential that a four-vector
parameter $\bmat s=(s^\alpha)$ should enter in variational
equations of the third order to make them obey the
pseudo-Euclidean symmetry. This parameter does not undergo any
variations.  Physically, it is responsible for   an intrinsic
dipole momentum of a relativistic test particle. As the problem
was posed on contact manifold, we obtain the solution in terms of
the coordinates on the contact manifold $C^3(1,M)$:
{ \def\k{\hbox{\ssb k}} \def\z{\hbox{\ssb z}} \def\E{\hbox{\ssb e}}
\begin{eqnarray}
\lefteqn{ \hbox{\ssb E}\;=\; \frac{\hbox{\ssb
v}^{{{\prime}}{{\prime}}}\times(\hbox{\ssb s}
-s_{\scriptscriptstyle0}\hbox{\ssb v})} {\bigl[(1+\hbox{\ssb
v}\q)(s_{\scriptscriptstyle0}{}^{2} +\hbox{\ssb
s}\q)-(s_{\scriptscriptstyle0}+\hbox{\ssb s}{{\bcdot}}\hbox{\ssb
v})^{2}\bigr]^{3/2}} }\nonumber\\ &&\hfill
\;-\;3\,\frac{(s_{\scriptscriptstyle0}{}^{2}+\hbox{\ssb s}\q)
\;\hbox{\ssb v}^{{\prime}}\!{{\bcdot}}\hbox{\ssb
v}-(s_{\scriptscriptstyle0}+\hbox{\ssb s}{{\bcdot}}\hbox{\ssb
v})\;\hbox{\ssb s}{{\bcdot}}\hbox{\ssb v}^{{\prime}}  }
{\bigl[(1+\hbox{\ssb v}\q)(s_{\scriptscriptstyle0}{}^{2}
+\hbox{\ssb s}\q)-(s_{\scriptscriptstyle0}+\hbox{\ssb
s}{{\bcdot}}\hbox{\ssb v})^{2}\bigr]^{5/2} } \;\hbox{\ssb
v}^{{\prime}}\times(\hbox{\ssb
s}-s_{\scriptscriptstyle0}\hbox{\ssb v}) \hfill\nonumber\\
&&\hfill\label{v-eq} +\;m\,\frac{(1+\hbox{\ssb v}\q)\,\hbox{\ssb
v}^{{\prime}} -(\hbox{\ssb v}^{{\prime}}\!{{\bcdot}}\hbox{\ssb
v})\,\hbox{\ssb v}} {(1+\hbox{\ssb
v}\q)^{3/2}(s_{\scriptscriptstyle0}{}^{2}+\hbox{\ssb
s}\q)^{3/2}}\;,
\end{eqnarray}
produced by any of the following Lagrange
functions,
\begin{eqnarray*}
L_{(i)}&=&\b\frac{s_{\scriptscriptstyle0}}{{s_{\scriptscriptstyle0}}^{2}+S\q}\cdot\frac{({s_{\scriptscriptstyle0}}^2+{\k_{(i)}}\q)(s_i-s_{\scriptscriptstyle0}v_i)
-s_i({\k_{(i)}}\bcdot{\z_{(i)}})}{({s_{\scriptscriptstyle0}}^2+{\k_{(i)}}\q)\,{\z_{(i)}}\q-({\k_{(i)}}\bcdot{\z_{(i)}})^2}
\cdot\frac{[W,(S-s_{\scriptscriptstyle0} V),{\E}_{(i)}]}{(S-s_{\scriptscriptstyle0} V)\q+(S\times
V)\q}\e\\ &&\b{}-\frac{m}{({s_{\scriptscriptstyle0}}^2+S\q)^{3/2}}\sqrt{1+V\q},
\e
\end{eqnarray*}
where some shortcut notations were introduced:
$$
\b
\k_{(i)}=S-s_{\mathstrut i}\,{\E}_{(i)},\qquad \z_{(i)}=(S-s_{\scriptscriptstyle0}
V)-(s_{\mathstrut i}-s_{\scriptscriptstyle0} v_{\mathstrut i})\,{\E}_{(i)}, \e $$
and
vectors ${\E}_{(i)}$ form a basis in $\R^3$ }

\begin{rem}{\rm By virtue of a certain proposition of \cite{Thesis} it is
not realistic to try to find any third order variational equation
with pseudo-Euclidean symmetry in four-dimensional space without
introducing into it some additional quantities, constructed from
the representations of the pseudo-Euclidean group.}
\end{rem}

Proposition
(\ref{P1})
immediately allows us to build the parameter-homogeneous form of the expression
(\ref{v-eq})
by means of the following prescription: if
$$
\dt\L={\cal E}_\alpha dx^\alpha\qquad\mbox{and}\qquad
\d(Ldt)=E_idx^i\otimes dt,
$$
then
$$
{\cal E}_\alpha
dx^\alpha=-\,\frac{dx^i}{d\tau}\cdot(E_i\circ\wp_\phi)\,dt+\frac{dt}{d\tau}\cdot(E_i\circ\wp_\phi)\,dx^i.
$$
So for (\ref{v-eq}) we obtain:
\begin{equation}\label{u-eq}
{\bmat{\cal E}}=
\frac{\ast\,{\bf\ddot{\bmat u}}\wedge\bmat u\wedge\bmat s}
{\|\bmat s\wedge\bmat u\|^3}
-3\,\frac{\ast\,{\bf\dot{\bmat u}}\wedge\bmat u\wedge\bmat s}
{\|\bmat s\wedge\bmat u\|^5}\,({\bf\dot{\bmat u}}\wedge\bmat s){{\bcdot}}(\bmat u\wedge\bmat s)
+\frac m{\|{\bmat s}\|^3}\left[\frac{{\bf\dot{\bmat u}}}{\|\bmat u\|}
\,-\,\frac{{\bf\dot{\bmat u}}{{\bcdot}}\bmat u}{\|\bmat u\|^3}\,\bmat u
\right]=0\,,
\end{equation}
and again Proposition
(\ref{P1})
helps to guess the family of four
Lagrange functions, each of which produces equation
(\ref{u-eq}):
\begin{equation}
\label{Lagr} \L_{(\alpha)}=\frac{\ast\,{\bf\dot{\bmat
u}}\wedge\bmat u\wedge\bmat s \wedge\bmat e_{(\alpha)}}{\|\bmat
s\|^2\|\bmat s\wedge\bmat u\|}\cdot\frac{ \bmat s\q
u_\alpha+(\bmat s\bcdot\bmat u)\,s_\alpha}{(u_\alpha \bmat
s-s_\alpha\bmat u)\q-(\bmat s\wedge\bmat u)\q}-\frac{m}{\|\bmat
s\|^3}\|\bmat u\|,
\end{equation}
with vectors ${\bmat e}_{(\alpha)}$ constituting a basis in $M$.
Equation (\ref{u-eq}) possesses the first integral
\begin{equation}
\label{firstint}
\frac{\bmat s\bcdot\bmat u}{\|\bmat u\|},
\end{equation}
and by comparison with (\ref{DAN}) and (\ref{Pir2}) we calculate that every
time we choose
\begin{equation}
\label{Pir3}
\bmat s\bcdot\bmat u=0,
\end{equation}
it describes the free motion of a relativistic top.
\section{Autoparallel reparametrization of geodesic curves}
It was argued in \cite{DGA95} that an arbitrary third order equation
$\bmat\xi:T^2M\to T^3M$ of the local form
\begin{equation}
\label{Solved}
\ddot u^\alpha=\xi^\alpha(\dot{u}^\beta,u^\beta,x^\beta)
\end{equation}
defines an autoparallel curve only in the case, when the functions
$\xi^\alpha$ satisfy (in terms of the vector field $\bmat \xi$)
the following commutation relations with the Liouville fields
(\ref{Zeta}):
\begin{equation}
\label{Tp}
\left\{
\begin{array}
{r@{=}l}
(T\wp)[\bmat\zeta^1,\bmat\xi]&(T\wp)\bmat\xi\\
(T\wp)[\bmat\zeta^2,\bmat\xi]&{\bf0},
\end{array}\right.
\end{equation}
which might be put into the local form by the following PDE system with
constant Lagrange multipliers $\mu$ and $\kappa$
\begin{eqnarray}
\dot{u}^\alpha-\frac13\frac{\partial\xi^\alpha}{\partial\dot
u^\beta}u^\beta&=&\kappa u^\alpha\label{Const1}\\
\xi^\alpha-\frac13\frac{\partial\xi^\alpha}{\partial u^\beta}u^\beta
-\frac23\frac{\partial\xi^\alpha}{\partial \dot{u}^\beta}\dot{u}^\beta
&=&\mu u^\alpha.\label{Const2}
\end{eqnarray}
It remains to solve equations (\ref{Const1}, \ref{Const2}),
and to find the functions $\xi^\alpha$ for the representation
(\ref{Solved}) of equation (\ref{u-eq}). In order to cast
equation (\ref{u-eq}) in the form (\ref{Solved}), solved with
respect to the highest order derivatives, we add to it one more
equation of general type
\begin{equation}
\label{Add}
{\bf\ddot{\bmat u}}\bcdot{\bmat u}=\|{\bmat u}\|^2\Psi({\bf\dot{\bmat
u}},\bmat u),
\end{equation}
and that will prescribe some kind of parametrization along the
unparametrized curves -- the solutions of (\ref{u-eq}). Next we
also make use of physical constraint (\ref{Pir3}). To proceed
further, contract vector equation (\ref{u-eq}) with the tensor
$*\,{\bmat u}\wedge{\bmat s}$ and differentiate first integral
(\ref{firstint}) twice. This helps to solve equation
(\ref{u-eq}) with respect to ${\bf\ddot{\bmat u}}$:
\begin{equation}
\label{Solved1}
{\bf\ddot{\bmat u}}=3\frac{{\bf\dot{\bmat u}}\bcdot\bmat u}{\|\bmat
u\|^2}\,{\bf\dot{\bmat u}}-3\frac{({\bf\dot{\bmat u}}\bcdot\bmat
u)^2}{\|\bmat u\|^4}\,\bmat u - m\,\frac{\|\bmat u\wedge\bmat s\|}{\|\bmat
s\|^3\|\bmat u\|}\,\ast\,{\bf\dot{\bmat u}}\wedge\bmat u\wedge\bmat
s+\bmat u\Psi.
\end{equation}
 Comparing (\ref{Solved1}) with (\ref{Solved}), we
rewrite (\ref{Const1}, \ref{Const2}) in terms of $\Psi$, and then
applying the compatibility conditions to the system of PDE
\{(\ref{Const1}), (\ref{Const2})\} shows that $\kappa=0$.  The
Ansatz for $\Psi$ is
$$
\Psi=\frac3{\|\bmat u\|^2}\left(\frac12\|{\bf\dot{\bmat u}}\|^2+\psi\right),
$$
and from (\ref{Const1}) there arises a constraint on possible functions
$\psi$:
\begin{equation}
\label{psi}
\bmat u\bld\frac{\partial\psi}{\partial{\bf\dot{\bmat u}}}=0.
\end{equation}
Let us apply symmetry concept to equation (\ref{Solved}). The
group of transformations of $M$ must not operate on the parameter
$\tau$. In case of pseudo-Euclidean group the generators read:
\begin{equation}
\label{Gen}
X=\Omega^{\alpha\beta} u_\alpha\frac\partial{\partial
u^\beta}+\Omega^{\alpha\beta} \dot{u}_\alpha\frac\partial{\partial \dot
u^\beta} +\Omega^{\alpha\beta} \ddot u_\alpha\frac\partial{\partial \ddot
u^\beta}+\Omega^{\alpha\beta} s_\alpha\frac\partial{\partial
s^\beta},
\end{equation}
with arbitrary skewsymmetric matrix parameter
$\Omega^{\alpha\beta}$. Now apply $X$ to equation
(\ref{Solved1}) and observe that if $\bmat a$ is a vector, then
$Xa^\alpha=-\eta^{\alpha\beta}\Omega_{\beta\gamma}a^\gamma$, where
$\eta^{\alpha\beta}$ is the constant canonical diagonal metric
tensor of pseudo-Euclidean $M$. This observation together with
(\ref{psi}) and (\ref{Const2}) suggests the solution
$$ \Psi=\frac3{\|\bmat u\|^2}\left(\frac12\|{\bf\dot{\bmat
u}}\|^2+A\|{\bf\dot{\bmat u}}\wedge\bmat
u\|^{4/3}\right),\qquad\mu=0, $$
with arbitrary scalar constant
$A$
\begin{prop}
The autoparallel curves in four-dimensional pseudo-Euclidean space describe the
motion of the free relativistic top and satisfy the equation
$$
{\bf\ddot{\bmat u}}=3\frac{{\bf\dot{\bmat u}}\bcdot\bmat u}{\|\bmat
u\|^2}\,{\bf\dot{\bmat u}}-3
\left[\frac{({\bf\dot{\bmat u}}\bcdot\bmat u)^2}{\|\bmat
u\|^4}-\frac12\frac{\|{\bf\dot{\bmat u}}\|^2}{\|\bmat
u\|^2}-A\frac{\|{\bf\dot{\bmat u}}\wedge\bmat
u\|^{4/3}}{\|\bmat
u\|^2}\right]\bmat u
- m\,\frac{\|\bmat u\wedge\bmat s\|}
{\|\bmat s\|^3\|\bmat u\|}\,\ast\,{\bf\dot{\bmat u}}\wedge\bmat
u\wedge\bmat s.
$$
The world lines are those among the extremal curves of the Lagrange function
(\ref{Lagr}), who agree with the physical constraint (\ref{Pir2}) \end{prop}
The constant $A$ corresponds to different ways of the parametrization of
world lines. One may chose $A=0$.

\end{document}